\documentstyle[epsfig]{l-aa}
\begin{document}
\thesaurus{02.04.01, 08.14.1, 08.16.6, 02.07.01 }
\title{Rapid differential rotation of protoneutron stars
       and constraints on radio pulsars periods}

\author{J-O.~Goussard\inst{2} \and P.~Haensel\inst{1,2} 
\and J.L.~Zdunik\inst{1}}

\institute{N. Copernicus Astronomical Center, Polish
           Academy of Sciences, Bartycka 18, PL-00-716 Warszawa, Poland
\and
D{\'e}partement d'Astrophysique Relativiste et de Cosmologie, 
     UPR 176 du CNRS, Observatoire de Paris, 
Section de Meudon, 
 F-92195 Meudon Cedex, France \\
{\em e-mail : goussard@obspm.fr, haensel@camk.edu.pl, jlz@camk.edu.pl} }
\offprints{J-O. Goussard}
\date{}
\maketitle
%
\begin{abstract}
Models of differentially rotating protoneutron stars are calculated, 
using realistic equations of state of dense hot matter. Various 
conditions within the stellar interior, corresponding to different 
stages of protoneutron star evolution, are considered. 
Numerical calculations are performed within the approximation 
of stationary equilibrium, using general relativistic equations 
of stationary motion of differentially rotating, axially symmetric stars
and using a numerical code based on spectral methods. 
 Families of differentially rotating models  
 of a given baryon mass  are calculated, using a 
two-parameter formula describing the angular velocity profile within 
a rotating protoneutron star. Apart from the usual 
``mass shedding limit'',  we introduce an additional 
``minimal mass limit'' for 
differentially rotating protoneutron stars resulting from a type II  
supernovae. 
Maximum angular momentum, which can be accommodated by 
 a protoneutron star within these limits 
is calculated, for various thermal conditions in stellar interior, for  
 a  baryon mass of $1.5~{\rm M}_\odot$. In the case  
 of  a thermally homogeneous (isentropic or isothermal) 
neutrino-opaque interior this maximum 
angular momentum turns out to be somewhat higher than that of 
 a cold neutron star of the same baryon mass, rotating uniformly 
at the mass shedding angular velocity.  However, if the protoneutron 
star has a thermal structure characteristic of   initial  
state, with  a low entropy (unshocked) core, and a high entropy 
(shocked)  outer half of baryon mass, the maximum angular momentum 
is significantly lower. This leads to a minimum period of uniform 
rotation of cold neutron stars of baryon mass $\sim 1.5~{\rm M}_\odot$, 
formed directly (i.e. without a subsequent significant accretion of mass) 
from protoneutron stars with shocked envelope, of about 1.7~ms and
strengthens the hypothesis that millisecond pulsars are accretion 
accelerated neutron stars.

\end{abstract}
%

\section{Introduction }
Newly born neutron stars are expected to be quite 
different from ordinary neutron stars, which 
 are a commonly accepted model of radio pulsars. 
Just after their formation in a gravitational collapse 
of a massive stellar core, a future neutron star 
is a hot (internal temperature $\sim 10^{11}~$K), 
lepton rich (one electron for  three nucleons) 
object - in view of these specific features it 
is called a {\it protoneutron star}. After some 
seconds, a protoneutron star transforms into 
a neutron star, loosing the lepton number excess 
via emission of neutrinos trapped in dense, hot 
stellar interior. The short period of neutrino burst 
is of paramount importance for the fate of collapsing 
star. 
 It is commonly accepted, on the basis of numerical 
 simulations of gravitational collapse of stellar cores, 
 that the prompt-shock mechanism  fails to make a 
 successful explosion corresponding to a SN II. 
 It is currently believed, that neutrino flow from a 
 protoneutron star can revive the stalled shock. It is 
 thus clear that the structure and evolution of a protoneutron star, 
 formed as a product of gravitational collapse, are crucial for 
 the success (or failure) of the accretion shock revival, and 
 therefore, for a successful SN II explosion.

While the lifetime of a protoneutron star 
(seconds) is 
negligibly short compared with the lifetime 
of a neutron star at the radio pulsar stage, 
it is some three orders of magnitude longer 
than the dynamical timescale for these objects 
(milliseconds). In view of this one 
can study the bulk evolution of protoneutron stars 
in the quasistationary approximation, treating convection just 
as an additional transport process in the stellar interior. 
Static properties of protoneutron stars, under various 
assumptions concerning their composition 
and equation of state  of hot, dense interior, 
were studied by numerous authors 
(Burrows \& Lattimer 1986, Takatsuka 1995, 
 Bombaci et al. 1995, Bombaci 1996, Prakash et al. 1997, 
Gondek et al. 1997). 

The formation of a protoneutron star (PNS) occurs on a
dynamical timescale. It involves compression, with an overshoot of 
central density, and a hydrodynamical bounce. PNS is expected to 
 begin its life pulsating around its quasistatic equilibrium. 
Radial pulsations of PNS were recently studied in (Gondek et 
al. 1997). 

 If the presupernova core was rotating  - even very slowly, 
the PNS is expected to be rather rapidly rotating. This is  
an inevitable consequence  of the conservation of the angular 
momentum of the collapsing core, which during a fraction of a 
second shrinks to about one percent of its initial radius. 
 The effects of rotation on the process of collapse of massive 
stellar core were studied in M\"onchmeyer \& M\"uller (1989),
Janka \& M\"onchmeyer (1989) and Yamada \& Sato (1994). 

Rapid rotation can have a strong effect on the structure of 
a PNS. The effect of rotation is stronger than in the case 
of cold neutron stars (NS), because  a hot PNS, due to the 
thermal and neutrino trapping effects,  is a more extended 
 object than a cold configuration of the same baryon (rest) mass.  
 Rapid uniform rotation of PNS was studied recently in 
(Hashimoto et al. 1995, Goussard et al. 1997), who considered 
also the consequences of the rotational properties of PNS for 
cold NS, into which a PNS transforms due to deleptonization 
and cooling. 

Uniform rotation of PNS should be considered as a first 
step in the study of their rotational properties. Even if 
the presupernova core was uniformly rotating, collapse 
would generate a significant amount of differential rotation 
 of PNS (during collapse, each collapsing annulus 
conserves -- to a very good approximation --  its angular momentum 
with respect to the rotation axis). Numerical simulations of 
collapse of rotating stellar cores indicate a characteristic 
 distribution of angular velocity within a newly formed 
PNS :  angular velocity of the matter, which is approximatively constant 
on coaxial cylinders, decreases with increasing 
distance from the rotation axis (M\"onchmeyer \& M\"uller 1989, 
Janka \& M\"onchmeyer 1989).

Evolutionary timescales of a PNS are at most seconds;  
actually, the lifetime of a PNS is shorter 
than about  twenty seconds (the time needed 
for the hot stellar interior to deleptonize).  
 The only mechanisms which we could  contemplate as the means 
of a sufficiently rapid transport of angular moment on such a short 
timescale are related to the superstrong magnetic fields, convection and, 
maybe, turbulent viscosity.   
 Our knowledge of the quantitative role of these effects, which 
could rigidify rotation of PNS, is at present insufficient  to make 
 reliable estimate of rigidification timescale. On the other hand, 
the viscous timescale, related to the transport of neutrinos within 
the hot, neutrino opaque interior of a PNS, is much longer than the 
PNS lifetime.

In the present paper we generalize our previous calculations of the models 
  of rapidly rotating PNS (Goussard et al. 1997) to the case of 
 differential rotation. In particular, we study to what extent the 
presence of differential rotation modifies the conclusions obtained 
assuming uniform rotation of PNS, referring to the limitations imposed 
on the periods of solitary pulsars formed directly from PNS. 
 Existing calculations of rapid differential 
rotation were restricted to the case of {\it cold} NS 
(Wilson  1972, Komatsu et al. 1989), and used 
schematic (unrealistic) 
equations of state of $T=0$  neutron star matter.  
 Our models of 
hot, differentially rotating PNS are constructed using a realistic 
EOS of dense hot matter, taking into account both the effects of 
high temperature, and of trapped neutrinos.

The model of the dense, hot interior of PNS is described in Sect. 2. 
We begin with a short description of the assumed thermal and 
lepton structure. Then we present the equation of state of the 
PNS matter, corresponding to various stages of the PNS evolution. 
 We discuss also characteristic timescales, relevant for PNS,  
and justify 
the assumption of stationarity (hydrostatic equilibrium) 
of rotating PNS. The mathematical formulation of the 
problem of stationary differential rotation of PNS 
is presented in Sect. 3, where we also consider the problem of 
stability of differentially rotating PNS. 
 The numerical method of solution 
of equations of stationary motion is described in 
Sect. 4, where we also discuss 
in some detail the problem of precision 
of  our  numerical solutions. 
 In Sect. 5 we present our models of rapidly, differentially rotating 
 PNS, and discuss in particular the mass shedding limits for these models, 
 at various stages of PNS  evolution. We study also  the relation 
 between differentially rotating PNS and uniformly rotating cold NS. 
 Finally, Sect. 6  
contains the discussion of our results and our conclusions.      
%
\section{Physical conditions  in the interior of a protoneutron star}
%
Our models of PNS are based on available results of (mostly 1-D) 
simulations of neutron star birth (Burrows \& Lattimer 1986, 
Janka \& M\"onchmeyer 1989, M\"onchmeyer \& M\"uller 1989, 
Burrows et al. 1995).  
We consider PNS just after it settled into a quasi-stationary state 
(we neglect pulsations which could be excited due to hydrodynamical 
overshoot and bounce). As we restrict ourselves to the period after 
the successful revival of the standing (accretion) shock, we can 
neglect the effects of accretion. In what follows, we will consider 
three situations. The first one corresponds to a very initial state 
of a PNS, which is then composed of a hot shocked envelope, 
with entropy per baryon (in the units of 
$k_{\rm B}$) $s_{\rm env}\sim 5-10$ up to some baryonic density $n_{\rm env}$
, and an unshocked core, characterized 
by $s_{\rm core}\sim 1$ above some density $n_{\rm core} > n_{\rm env}$. 
 The layer with  $n_{\rm env}<n<n_{\rm core}$ 
constitutes a transition region between the unshocked core and 
the shocked envelope. 
Our standard model of a  PNS will have  
baryon (rest) mass of $M_{\rm bar}=1.5~{\rm M}_\odot$. Then, the shocked envelope 
will contain about $0.7~{\rm M}_\odot$ of the baryon mass. 
Both core and envelope are neutrino opaque - a neutrino transparent outer 
layer of the envelope has a negligible mass, and was not included. 
We will hereafter call such a protoneutron star configuration an early-type
PNS (EPNS).

The second situation 
 (established after  0.2-0.5 s, see Burrows \& Lattimer 1986, 
Keil et al. 1996) 
will correspond to a thermally homogeneous  PNS. 
  At this stage, the thermal state of 
PNS interior (below the neutrinosphere)  is characterized by $s\sim 2$, 
and some characteristic lepton fraction $Y_l$. This situation will be referred 
to as a late-type PNS (LPNS).

Further cooling 
and deleptonization transforms PNS into a NS of the same $M_{\rm bar}$. 
 The structure of a sufficiently cool NS 
(internal temperature $T<10^{10}$~K)  
 is determined 
 essentially  by the $T=0$ EOS; due to strong degeneracy  thermal effects 
in the EOS of the dense NS interior can be completely neglected. This
will correspond to a third situation considered by us.  
The three situations, described in this paragraph, correspond to 
quite different EOS of the stellar interior, described below.
\subsection{Equations of state and stationary models of PNS}
We will assume a well defined ``neutrinosphere'' which 
separates a hot, neutrino-opaque interior from colder 
 neutrino--transparent envelope. 
Hot dense matter  in 
the hot, neutrino opaque interior is composed 
of nucleons (both free and bound in nuclei) and 
leptons (electrons, positrons,  and neutrinos and antineutrinos 
of all flavors; for simplicity we do not 
include muons). Nucleon component of hot matter will be 
described by one of the models developed by Lattimer 
\& Swesty (1991), corresponding to  the incompressibility 
of symmetric nuclear matter at saturation density $K=220~$MeV. 
 The local state of matter 
in the hot interior is determined by the baryon (nucleon) number 
density $n$, the net electron fraction $Y_e=(n_{e^-}-n_{e^+})/n$, 
and the net fraction of trapped electron-neutrinos, $Y_\nu 
= (n_{\nu_e}-n_{\bar\nu_e})/n$. 
 An initial stage  
of PNS evolution is characterized by a significant trapped lepton number 
$Y_l=Y_\nu  +  Y_e \simeq 0.4$. Under these conditions, both 
 electrons and electron--neutrinos are strongly degenerate. 
 The deleptonization, due to the 
$\nu_e$ diffusion outward (driven by the gradient of the 
chemical potential of $\nu_e$)  and  subsequent emission from their 
neutrinosphere, implies a strong decrease of $Y_l$ on a timescale 
of seconds (Sawyer \& Soni 1979, Prakash et al. 1997). 
 After the deleptonization (which for simplicity is assumed to be 
complete, with $Y_\nu=0$), there is no trapped lepton number, 
 so that neutrinos trapped within the hot interior do not 
influence the beta equilibrium of nucleons, electrons and 
positrons. The neutrino diffusion outward (towards the 
neutrinosphere) is then driven by the temperature 
gradient. 
 The neutrinosphere is  determined by the condition that 
optical thickness of the layer above it, 
for electron neutrinos  with mean energy 
corresponding to $T$ at the neutrinosphere, be unity. 
\begin{figure}     
\centerline{
\epsfig{figure=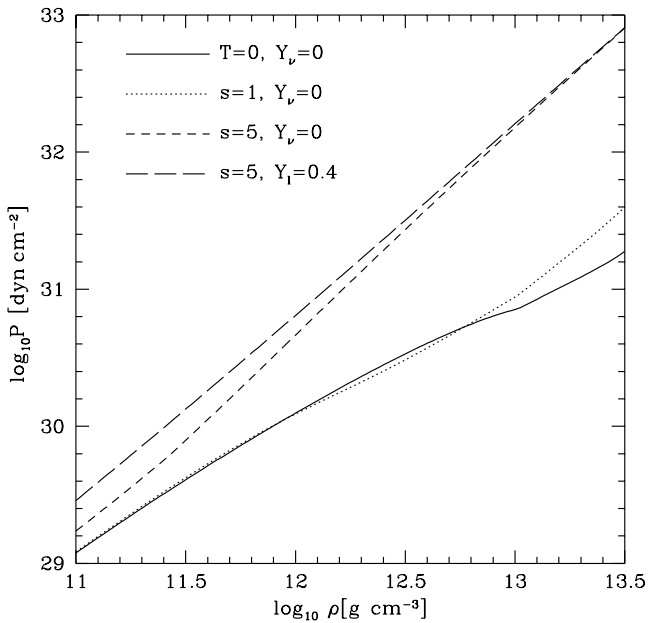,width=8.7cm}}
\caption[]{
Pressure versus matter density in the shocked ($s=5$) envelope of a PNS 
(dashed and long-dashed lines). Two curves correspond to two limiting 
cases of trapped lepton number: long dashes - maximal trapped lepton 
fraction $Y_l=0.4$; short dashes - no trapped lepton fraction 
($Y_\nu=0$).  For the sake of comparison, we show also the EOS in 
the standard NS case of cold matter (solid line, $T=0$), and that 
corresponding to low entropy  and no trapped lepton number 
(dotted line, $s=1$, $Y_\nu=0$).  
\label{fig:EOSenv}}
\end{figure}

The effect of high $s$ on the EOS of the shocked envelope of 
a PNS can be seen by comparing the pressure-mass density 
dependence in the low  and high entropy  cases (Fig. 1). 
For $s=5$ the pressure  
increases by more than an order of magnitude at 
$\rho=10^{13}~{\rm g~cm^{-3}}$, as compared to the case with 
$s=1$; the relative effect of $s$ on $p$ decreases with decreasing 
density. Above $10^{12}~{\rm g~cm^{-3}}$, the effect of trapped 
lepton number on the EOS of 
shocked envelope with $s=5$ is much smaller than 
that of $s$. 

Actually, our models of EPNS were constructed only with a $Y_\nu=0$
shocked envelope. This was done mostly for the sake of simplicity, and
also to counterbalance some oversimplifications of our model with 
constant trapped lepton fraction, which would clearly overestimate the 
role of neutrinos (in numerical simulations, $Y_l$ decreases outward,
c.f., Keil et al. 1996). However, we think that a more detailed treatment
of the shocked envelope would not change our main conclusions. On one hand,
the layer with $\rho < 10^{12}~{\rm g~cm^{-3}}$, in which the assumption 
of $Y_\nu=0$ could lead to a significant underestimating of pressure (Fig. 1),
contains a very small fraction of the stellar mass. On the other hand, the
inclusion of $Y_\nu >0$ would further stiffen the EOS of the shocked envelope,
and could only make our conclusions concerning the role of the shocked
envelope even stronger (see Sect. 5.1).
\subsection{Timescales  and stationarity}
The EOS of PNS evolves with time, due to the deleptonization, 
 which changes the composition of the matter, and due to 
temperature change. However, significant changes of 
both $T$ and $Y_l$  within the 
hot, neutrino opaque interior occur on a timescale 
of seconds, if only diffusive processes are operating. 
The evolutionary timescales can be shortened, if convective 
transport is included. Still, these evolutionary timescales 
 are orders of magnitude longer than the dynamical 
timescale, governing the  readjustment of pressure and 
gravity forces.  

Our treatment of PNS is based on the assumption, 
that convective motions do not influence directly the bulk
mechanical equilibrium of rotating PNS.  For the convective 
motions to be not important dynamically, convective matter 
velocities should be significantly smaller than the sound 
velocity  within the neutrino opaque  PNS interior. Velocity of sound, 
$v_{\rm s}$,  can be 
calculated from the EOS of the PNS matter using 
\begin{equation}
\left({v_{\rm s}\over c}\right)^2= 
\gamma{p\over e + p}~,
\label{v_s}
\end{equation}
where $e$ is the mass-energy density, and 
the adiabatic index $\gamma$ is defined by 
\begin{equation}
\gamma= {n \over p} \left({{\rm d}p\over {\rm d}n}\right)_{Y_l,s}~.
\label{gamma}
\end{equation}
In the case of cold NS matter the appropriate $\gamma$ is that calculated 
at a fixed $Y_e$ (Gondek et al. 1997).
\begin{figure}     
\centerline{
\epsfig{figure=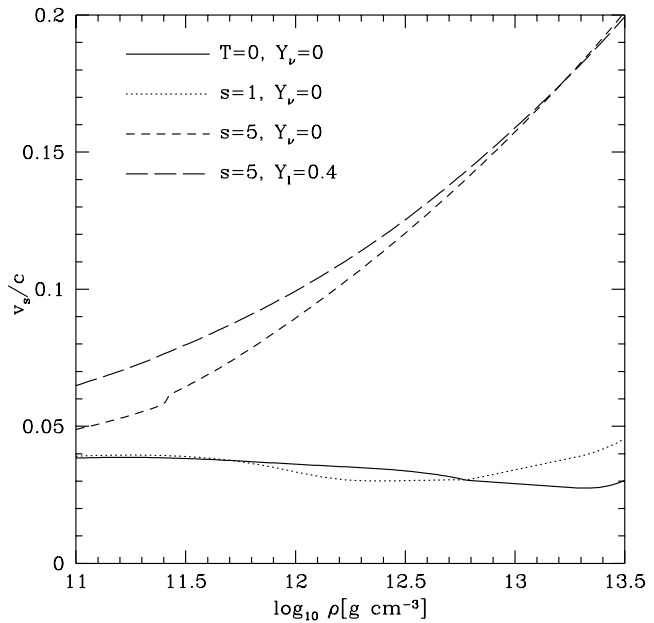,width=8.7cm}}
\caption[]{{\bf a.}
Velocity of sound (in the units of the velocity of light $c$) 
versus matter density, in the bloated envelope of PNS (dashed 
line and long-dashed line). The density is bounded from above 
by that corresponding to our assumed value of $n_{\rm env}$. 
The case of $Y_\nu=0$ (short dashed line) corresponds to no 
trapped lepton number, while that of $Y_l=0.4$ (long-dashed line) is 
that of maximum trapped lepton number. For the sake of comparison, 
we show also the case of cold matter (solid line, $T=0$), and 
that with low entropy and no trapped lepton number (dotted line, 
$s=1$, $Y_\nu=0$). 
\label{fig:vs1}}
\end{figure}
\addtocounter{figure}{-1}
\begin{figure}     
\centerline{
\epsfig{figure=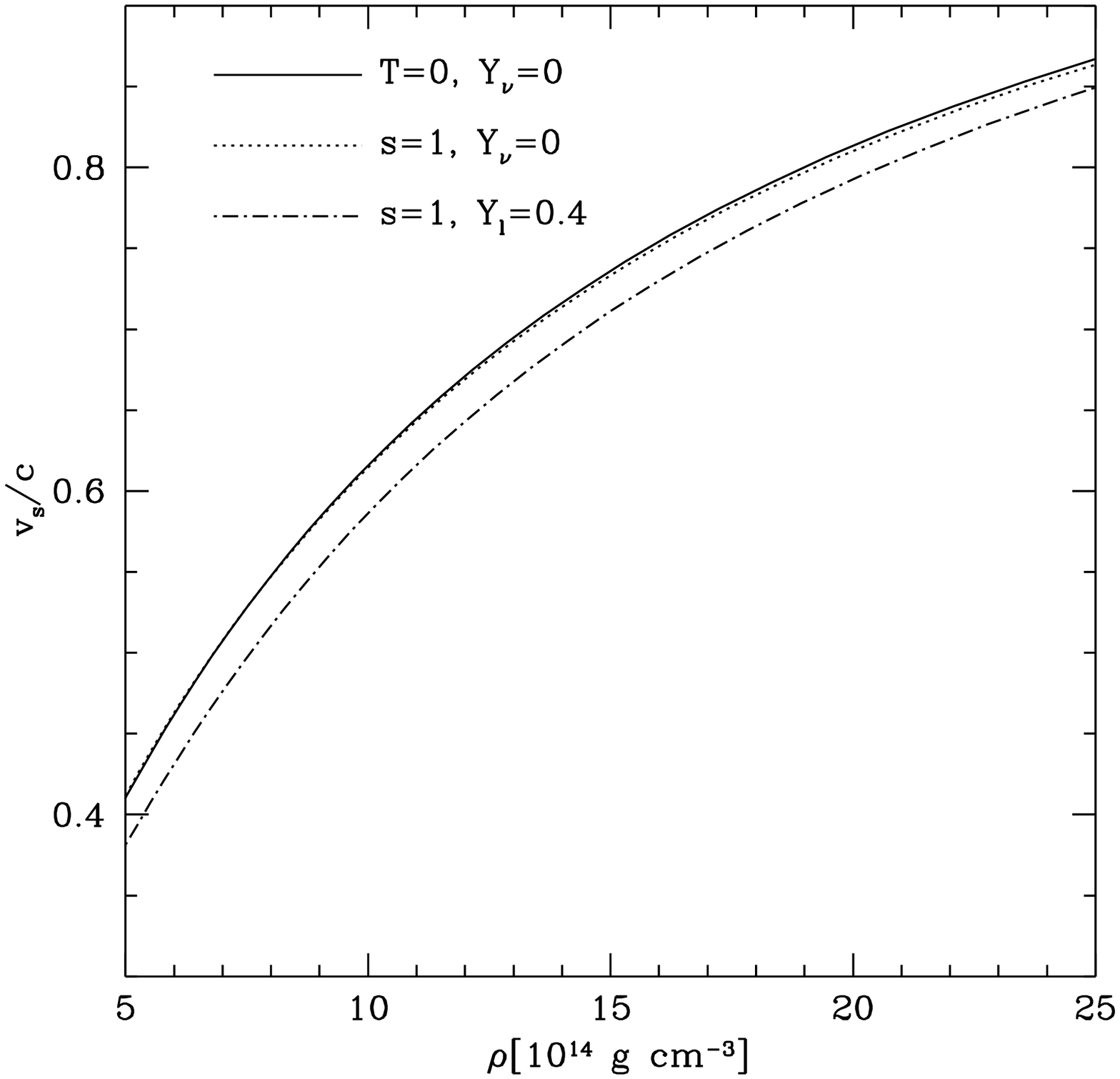,width=8.7cm}}
\caption[]{{\bf b.}
Velocity of sound (in the units of velocity of light $c$) 
versus matter density, within the protoneutron star core 
($n>n_{\rm core}$). Notation as in Fig. 2a. 
\label{fig:vs2}}
\end{figure}

In Fig.2a we plot the values of $v_{\rm s}/c$ versus matter density, 
within the shocked envelope of a PNS, and compare these values with 
those corresponding to different physical conditions in this density 
interval. The sound velocity in the shocked envelope ($s=5$) is 
significantly higher than that corresponding to the  matter at 
$s=1$. Strongest effect of high $s$ is seen for 
$\rho>
10^{12}~{\rm g~cm^{-3}}$. Trapped lepton number increases $v_{\rm s}$, 
but its effect  is relatively small as compared to that of $s=5$ 
for $\rho>10^{12}~{\rm g~cm^{-3}}$. 
Sound velocity in the shocked envelope is significantly higher than 
the velocities of convective motions at the early stage of evolution 
of a protoneutron star obtained in numerical calculations (Keil et al. 1996). 
This conclusion is even stronger in the  case of the unshocked 
 core (Fig. 2b). The effect of finite entropy on the value of 
$v_{\rm s}$ is there negligible, because of the strong degeneracy of 
the matter, while trapped lepton number slightly decreases it.

A very important timescale is that of viscous dissipation of 
differential rotation. In the absence of convection and 
turbulence, shear viscosity is due to the diffusion 
of neutrinos. We will estimate the neutrino viscosity 
under the conditions prevailing in the dense hot core 
of PNS, using the general scheme of van den Horn \& 
van Weert (1981a,b, 1984) for the calculation 
of transport coefficients of neutrinos trapped 
in dense, hot matter, combined with  neutrino 
mean free paths as quoted in Burrows \& Lattimer 
(1986). We obtain three expressions, corresponding 
to three different physical regimes for dense 
hot matter with trapped neutrinos. The earliest 
epoch corresponds to strongly degenerate 
electron neutrinos. 
In the dense PNS core, nucleons are degenerate, 
and neutrino viscosity in this regime (referred to  as 
${\rm D}n,{\rm D}\nu$), 
resulting from transport of degenerate 
electron neutrinos, is  
\begin{eqnarray}
\eta_\nu({\rm D}n,{\rm D}\nu) 
&\simeq& 3.1\cdot 10^{23}~\rho_{14}\times \nonumber\\
&&  \times (Y_\nu/0.05)
(T/10~{\rm MeV})^{-2}~
{\rm g~cm^{-1}~s^{-1}},
\end{eqnarray} 
where $\rho_{14}=\rho/(10^{14}~{\rm g~cm^{-3}})$ 
(our formula is in good agreement with the results 
obtained by Goodwin \& Pethick (1982), who used the 
methods of the theory of Fermi liquids). 
 After deleptonization, neutrinos become   
non-degenerate, and the neutrino viscosity in 
degenerate nucleon gas becomes somewhat lower,
\begin{eqnarray}
\eta_\nu({\rm D}n,{\rm ND}\nu) &\simeq & 9.7\cdot 10^{22}\times \nonumber \\
&&~\times (T/{\rm 10~MeV})/\rho_{14}^{2/3}~
{\rm g~cm^{-1}~s^{-1}}. 
\end{eqnarray} 
Finally, at subnuclear densities we can have 
 non-de\-ge\-ne\-rate neu\-trinos trapped in a dense 
hot nucleon gas, which corresponds to  
\begin{eqnarray}
\eta_\nu({\rm ND}n,{\rm ND}\nu)
&\simeq &
2.3\cdot 10^{23} \times \nonumber \\
&&~\times (T/{\rm 10~MeV})^2
/\rho_{13}~
{\rm g~cm^{-1}~s^{-1}}. 
\end{eqnarray}
 In all cases the viscous timescales within the  
 stellar interior, 
$\tau_\nu = l^2\rho/\eta_\nu$, with $l\sim 10~{\rm km}$, are 
 much longer than the lifetime of a PNS. Both magnetic field 
 effects and local turbulence could shorten  viscous timescales, 
 but they are expected to be still much longer than the 
 dynamical timescale.
 
The above discussion shows, that evolutionary time\-scales with\-in  
 a PNS, resulting from the transport processes in the stellar 
interior,    are significantly longer than the dynamical timescale 
for PNS.  
  In view of this, we are able to decouple 
the PNS evolution from its dynamics. 
 We can then treat PNS rotation 
in a  stationary approximation, with a well defined EOS of 
the PNS matter. 

\section{Formulation of the problem}
The problem of the calculation of the stationary state of
differentially rotating cold neutron stars, within the framework
of general relativity, was considered previously by some authors (Komatsu et al. 1989,
Wilson 1972, 1973), who restricted themselves, however, 
to polytropic (and hence unrealistic) EOS of dense matter. 
 Here we will extend the methods used at $T=0$
to the case of hot PNS with realistic  EOS. 
We will use the notation and formalism
developed in Bonazzola et al. (1993), hereafter referred
to as BGSM.
%
\subsection{Equation of stationary motion}
As we already mentioned in paper I, 
one of the problems introduced by finite temperature
and differential rotation, if one does not make any further assumption
on the equilibrium configuration, is that the equation of stationary
motion does not have a first integral (Bardeen 1972).
This equation reads (eq. (5.8) of BGSM) :\\
\begin{equation}
\frac{\partial_{i} p}{e+p} + \partial_{i}\ln (\frac{N}{\Gamma}) = 
-F \partial_{i} \Omega~ , \label{statmot}
\end{equation}
where $p$ is the pressure, $e$ the mass-energy density, $N$ the lapse
function appearing in the space-time metric, $\Gamma$  is the Lorentz factor
due to rotation, $\Omega$ is the angular velocity and $F = u_{\phi} u^{t}$
where $u$ is the fluid 4-velocity.

This equation is the general relativistic equivalent of a well-known
Newtonian formula (see e.g. Tassoul 1978). It is straightforward to
show that a {\it sufficient} condition for (\ref{statmot}) to be
integrable is 
\begin{equation}
T = T(n)~.  \label{tden}
\end{equation}
Eq.(\ref{statmot}) can then be rewritten
\begin{equation}
\partial_{i}[ G + \ln (\frac{N}{\Gamma})] = 
-F \partial_{i} \Omega~ , 
\end{equation}
where the heat function $G$ is defined by
\begin{equation}
G = \int{\frac{dp}{e(n,T(n))+p(n,T(n))}}~,
\end{equation}
which is not to be confused in non-isentropic cases with
the logarithm of the dimensionless relativistic enthalpy per baryon
$H = \ln(f) = \ln[(e+p)/(n m_0 c^2)]$ (see paper I).
Then, as in BGSM (\S 5.1), we apply the Schwarz theorem and 
find that either 
\begin{equation}
\Omega = const.
\end{equation}
or
\begin{equation}
F = F(\Omega). \label{fom}
\end{equation}
 This corresponds to the well-known result that, assuming barotropic 
equilibrium ($T=T(n)$), 
the rotation of the star is either uniform or cylindrical. Being interested
in differential rotation, we restrict ourselves to the case (\ref{fom}).

The rotation law is obtained by solving 
Eq. (5.17) of BGSM (we correct here a misprint of this paper) :
\begin{equation}
F(\Omega) - \frac{A^4 B^2 r^2 \sin^{2}\theta(\Omega - N^{\phi})}
{N^2 - A^4 B^2 r^2 \sin^{2}\theta(\Omega - N^{\phi})^2} = 0, \label{solvf}
\end{equation}
where $A$, $B$ are metric potentials and $N^{\phi}$ is the shift vector
(see BGSM for a precise definition).

One then obtains the first integral of motion in the form :
\begin{equation}
G + \ln(\frac{N}{\Gamma}) + \int{F(\Omega)d\Omega} = const.
\label{firstint}
\end{equation}
This enables us to calculate $G$
in the whole star and thus the other thermodynamic quantities,
using the $G$-tabulated EOS of the form $e=e(G)$, $p=p(G)$, $s=s(G)$ 
(see \S 4.2).

Finally, in the case of general relativistic thermal equilibrium, one can
show, in the context of the Eckart (1940) or Carter (1983) theory of heat 
diffusion, that
\begin{equation}
T^{\star} = T \frac{N}{\Gamma} e^{\int{F d\Omega}} = const.
\end{equation}
where $T^{\star}$ is the redshifted temperature which would be measured
by a distant observer, 
corresponds to thermal  equilibrium (no heat flux). Such a $T$-profile leads to 
a simple form for the first integral of motion :
\begin{equation}
\mu^{\star} = \mu \frac{N}{\Gamma} e^{\int{F d\Omega}} = const.
\end{equation}
where $\mu$ is the baryon chemical potential. The $T$-profile is
then determined from 
\begin{equation}
\frac{\mu(n,T(n))}{T(n)} = const.
\end{equation}
which allows us to use the previous formulation (\ref{firstint}) of the 
first integral of motion.
\subsection{Differential rotation law}
There is obviously an infinite number of degrees of freedom in the choice
of the $F$-law. However, we wish to have a law which:  (1) approximately
reproduces the $\Omega$-profile obtained in 2D-simulations, i.e. essentially
decreases toward the edge of the star;  (2) is a  ``simple'' law, in
order to limit the number of free parameters.  
 We thus chose the
rotation law used by Komatsu et al. (1989) :
\begin{equation}
F(\Omega) = R_{0}^2 (\Omega_{c} - \Omega) \label{Fkomat}
\end{equation}
where $\Omega_{c}$ is the central angular velocity and $R_0^2$ is a
free parameter, with the dimension of the square of a length (denoted $A$ in
Komatsu et al. 1989).
This simple rotation law allows for a 
rapid solution of (\ref{solvf}) since it becomes then a 
cubic equation.

In the Newtonian limit,  the rotation law (\ref{Fkomat}) 
yields the angular velocity
\begin{equation}
\Omega = \frac{R_0^2 \Omega_c}{R_0^2 + r^2 \sin^{2}\theta}. \label{onew}
\end{equation}
We see that  $R_0$ is thus a characteristic scale of spatial 
variation of $\Omega$ : when
$R_0$ is small the rotation profile is very steep, leading to a constant 
specific angular momentum distribution,  and when it is 
large the rotation is almost uniform. Numerical experiments have shown 
that the general relativistic profile has the same qualitative 
properties as the Newtonian one. 
Note that this Newtonian profile is  
also the equivalent,  for cylindrical rotation,  to  the one used by
 M\"onchmeyer \& M\"uller (1989) and Janka \& M\"onchmeyer (1989) at the 
beginning of their calculations with spherical rotation. It is also the one used
by M\"uller and Eriguchi (1985) in their calculations
 of ``fizzling'' white dwarfs.
%
\subsection{Stability of differentially rotating PNS}
Differentially rotating self-gravitating bodies are subject to
various dynamical instabilities, which may deeply affect their structure
(see e.g. Zahn 1993). 
In what follows, we study to what extent our models of hot PNS
are stable regarding the two main axisymmetric instabilities due
to the differential character of rotation,  appearing in 
barotropic configurations : the Rayleigh and the shear instability.
We then briefly discuss the case of secular instabilities linked to
the spontaneous symmetry breaking phenomenon.

\subsubsection{Rayleigh instability} 
It is straightforward to show
that the newtonian $\Omega$-profile (\ref{onew}) satisfies the classical
Rayleigh stability criterion, i.e. :
\begin{equation}
\frac{{\rm d}j}{{\rm d}(r\sin\theta)} = 
\frac{{\rm d}(r^2\sin^2\theta\Omega)}{{\rm d}(r\sin\theta)} > 0,
\label{rayleigh}
\end{equation}
where $j$ is the specific angular momentum of baryonic matter. This
criterion yields also a {\it sufficient} condition for stability
in the case of cylindrical rotation laws, i.e.
\begin{equation}
\frac{{\rm d}j}{{\rm d}\Omega} < 0,
\end{equation}
and one can show that the general relativistic $\Omega$-profile
obtained with the law (\ref{Fkomat}) fulfils this last condition 
{\it in the isentropic case} (Komatsu et al., 1989, \S 3.1).
For the full general relativistic formulation in non-isentropic
cases, we checked that all our models verified (\ref{rayleigh}).
As this stability analysis is a local one in which the gravity does
not enter, we infer that (\ref{rayleigh}) is a good criterion even
in the general relativistic case. 
\subsubsection{Shear instability} 
The case of the shear instability is much more
complicated, and we will restrict ourselves to crude
estimates. Our configurations, in which the
Reynolds number (which compares turbulent to molecular viscosity)
is of the order of $Re =R^2 \rho\Omega/\eta \sim 
10^3$ (see Section 2.2), 
are very likely to be unstable (see Zahn 1993). 
In principle, this instability  could be blocked by a  stable entropy 
or specific angular momentum stratification inside the star.
However, even if the entropy and specific angular momentum
gradients of our models were all stable (apart from  a thin outer layer
of the star), these gradients 
are too weak to stabilize the flow. 
Indeed, the Richardson number $Ri$,
which compares the efficiency of the stabilizing $s$ and $j$-gradients 
to destabilizing effect of shear, can be written in the Newtonian case as 
\begin{eqnarray}
Ri &=& \left[\left(g 
- \frac{j^2}{r^3}\right)\left(\frac{1}{\gamma}\frac{{\rm d}\ln(p)}{{\rm d}r}
-\frac{{\rm d}\ln(\rho)}{{\rm d}r}\right) 
+ \frac{1}{r^3}\frac{{\rm d}j^2}{{\rm d}r}\right] \times \nonumber \\
&&\times \left(\frac{{\rm d}U}{{\rm d}r}\right)^{-2}
\end{eqnarray}
where $g$ is the gravitational acceleration, $\gamma$ the adiabatic index, 
$\rho$ the mass density and ${\rm d}U/{\rm d}r$ the vorticity.
In view of the fact,  that our models are almost adiabatic, i.e. that
\begin{equation}
\left\bracevert\frac{1}{\gamma}\frac{{\rm d}\ln(p)}{{\rm d}r}
-\frac{{\rm d}\ln(\rho)}{{\rm d}r}\right\bracevert < 0.5
\end{equation}
 in the whole star, and that as $R_0^2$
decreases the $j$-distribution tends to a $j = const.$ one (giving thus
a marginally Rayleigh-stable stratification), it was not a surprise to find
that almost all our models verified $Ri < Ri_{\rm crit}=\frac{1}{4}$ (at least in some
part of the star), which is a sufficient condition for instability.

However, the global effect of
the shear instability is to suppress the phenomenon that causes it,
namely, differential rotation. As the maximum  angular momentum
of PNS decreases with decreasing differential rotation (see \S
5), the introduction of this instability can only decrease the maximum 
angular velocity of such stars.
\subsection{Non-axisymmetric instabilities}

A rapidly rotating PNS can be affected by secular instabilities,
driven by the gravitational radiation reaction (GRR instability) or 
by viscous dissipation. However, detailed 
calculations performed for hot uniformly
rotating NS suggest that, for temperatures exceeding $10^{10} {\rm K}$, 
the GRR instability can only slightly decrease the maximum rotation frequency
(Cutler et al. 1990, Ipser \& Lindblom 1991, Lindblom 1995, 
Yoshida \& Eriguchi 1995, Zdunik 1996). If the instability is
driven by viscous dissipation, Bonazzola et al. (1995) have shown
for cold uniformly rotating NS
that only very stiff equations of state are likely to allow for 
spontaneous symmetry breaking. 

As our models are differentially rotating and hot, no previous
study applies to our EPNS and LPNS. To estimate to what extent 
our models could be plagued by spontaneous symmetry breaking, we
used the {\it newtonian} criterion of stability $T/|W| < 0.14$,
where $T$ is the kinetic energy and $W$ the gravitational energy
of the star. As our PNS models are slowly rotating compared to 
cold NS, we were not surprised to find that $T/|W| \ll 0.14$ for
the EPNS models. For LPNS, we found some models for which 
$T/|W| \sim 0.12-0.14$, which, considering the approximate 
criterion we used, cannot be viewed as conclusive of symmetry breaking.
Furthermore, inclusion of symmetry breaking effects can only
decrease the maximum angular momentum of PNS and thus, their maximal
rotational velocity, which would only strengthen our conclusions.

\section{Numerical method}
%
\subsection{Axisymmetric code}
We used the same code based on spectral methods as in paper I, modified 
to take into account
differential rotation. We refer the reader to our previous paper
for a brief description of the code and tests of 
precision, which  we made in the
case of uniform rotation. Here we briefly outline the different or 
new features of the code.

The global relative error, evaluated by the means of the virial
 identity test  (see BGSM), is $\sim 10^{-5}$. This is  by two orders
of magnitude better than the precision reached in paper I. 
 Such an increase of precision was possible 
 due to  new interpolation procedure used for the EOS (see
next section).

We used two grids in $r$ for the star and one in the  $1/r$-variable  for the
exterior, of $N_r$ Chebychev coefficients, with $N_r = 65$ in the interior 
and $N_r = 33$ in the exterior, and one
grid in $\theta$ with $N_\theta = 33$. We checked that taking  a
greater $N_r$ or $N_\theta$ does not change the results by more
than a few $10^{-6}$ at most, which stays within 
 the global precision reached. Let us stress that such a low number
of grid points is sufficient to reach high accuracy within a
spectral method (this would not be possible in the case of a finite difference
scheme). 

The temperature drop at the ``neutrinosphere'' is not always located on
the border of the internal grid in $r$, which could influence
the precision (remember that the spectral methods are very sensitive to 
discontinuities). We checked that, in fact, even when the 
 ``neutrinosphere'' is far from the border of the grid, the global
physical properties of the star did not change much for the late-type 
models of PNS (we found also relative variations of a few 
$10^{-6}$ at most). For early-type models of PNS, the results were more
sensitive to the location of the ``neutrinosphere'', 
and we were obliged to position the grids quite precisely
to reach the former precision.
 
%
\subsection{Interpolation procedure for the EOS}
The precision of the axisymmetric code is very sensitive to the 
thermodynamic consistency of the interpolated EOS. In particular, 
 the following  thermodynamical relation must be fulfilled to ensure a 
good precision of the code, 
\begin{equation}
\frac{{\rm d}p}{{\rm d}G} = e + p \label{eoscons}~. 
\end{equation}
If the above relation is  only approximately fulfilled  by the EOS (as is the case when
 various thermodynamical quantities are independently interpolated)
the equation of stationary motion cannot be strictly satisfied, neither.  
This influences greatly the value of the virial parameter and thus,
the precision of the calculations. 
In order to have a consistent EOS (in the sense that (\ref{eoscons}) is
verified),
we used an original procedure, adapted from a method derived by Swesty (1996). 

Once the temperature profile is chosen, the EOS can be tabulated.
The heat function $G$ is calculated by interpolating the
quantity $1/(e+p)$ in the integrand in Eq.(4)  
 as a function of $p$ using cubic splines and
integrating this interpolating polynomial.
 The quantity $G$ can then be chosen as the new 
interpolation variable : as this is
a quantity  given directly by the first integral of motion (see
\S 3) this ensures a rapid calculation of the thermodynamical
quantities. Futhermore, following Swesty, we use the fact that
${\rm d}p/{\rm d}G$ is known 
and interpolate $p$ and ${\rm d}p/{\rm d}G$ by means of a Hermite interpolation 
scheme of degree 3. The energy density is obtained at each point by derivating 
the interpolant and using (\ref{eoscons}). Other thermodynamical quantities
can be interpolated independently since they do not enter  this
relation. In doing so, we ensure that (\ref{eoscons}) is valid for the whole
interpolation polynomial and not only at the interpolation points.

With such an interpolation procedure, we were able to lower the virial
error parameter by two orders of magnitude.
\section{Numerical results}
%
\subsection{Static models of early-type PNS and ``minimal mass'' limit}
Early-type PNS, with a high-entropy, bloated envelope, are 
short living structures. Their lifetime might be as 
short as  100-200~ms 
(if one includes convective motions, see Keil et al. 1996), 
but this is still a long time 
on a dynamical timescale. 
As early-type models of PNS have not been investigated in paper I,
we study here the static properties of these objets. 
The gravitational and baryonic mass of static models versus central
baryonic density are displayed in Fig. \ref{fig:statmass}. Let us 
notice, that we have restricted ourselves to the range of masses, 
 consistent with assumed thermal structure of PNS. 
\begin{figure}     
\centerline{
\epsfig{figure=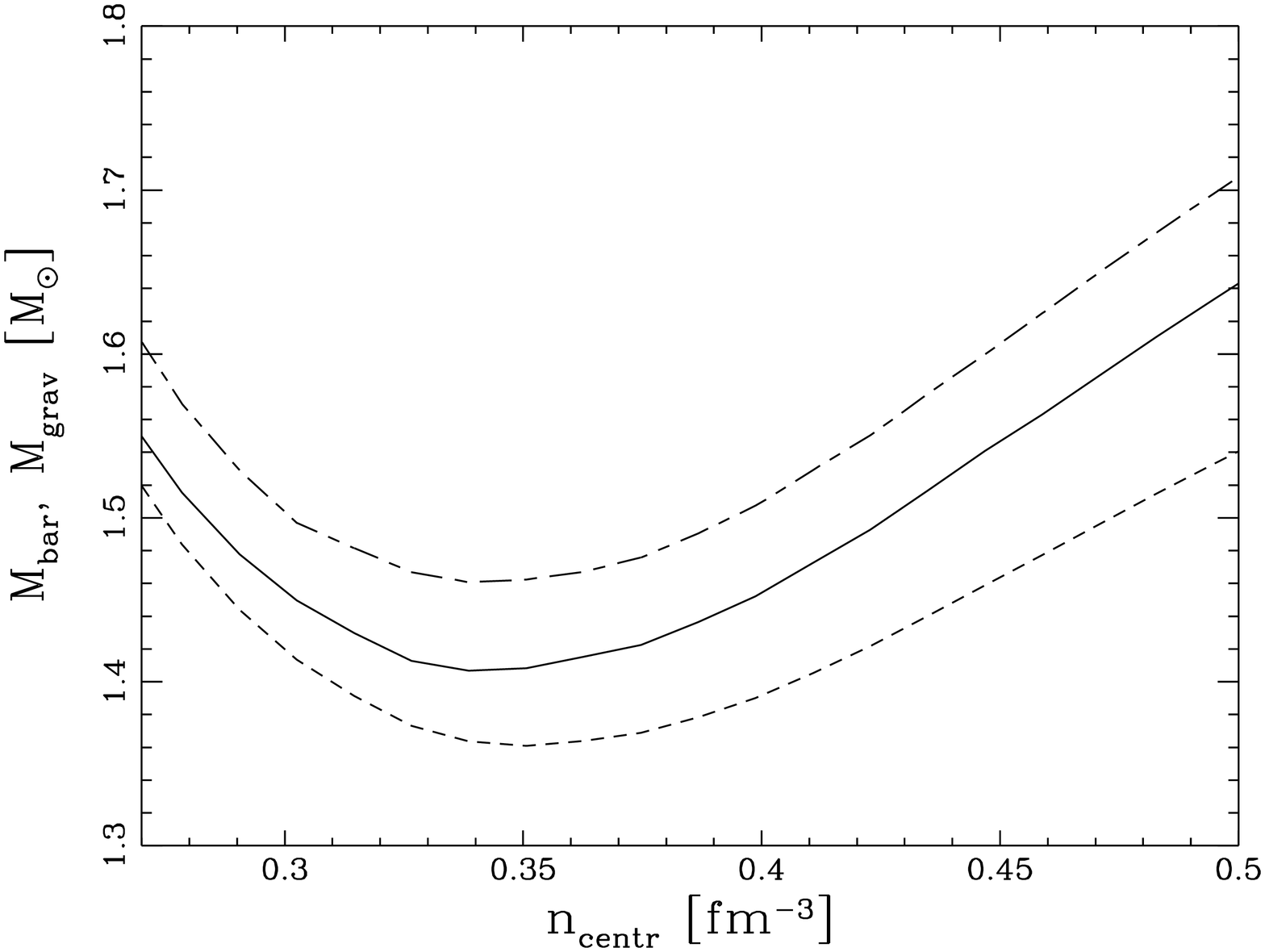,width=8.5cm,angle=-90}}
\caption[]{
Baryon and gravitational mass of static models of early-type PNS versus
central density, $n_{\rm centr}$. 
 The solid curve is for baryonic mass and the dashed curve 
for gravitational mass. The upper long-dash/short-dash curve is 
the baryon mass, $M_{\rm bar}$, 
calculated for 
differentially rotating configurations at $R_0^2 = 1~{\rm km^2}$ and 
$R \Omega_c = 0.284~{\rm c}$, where $R$ is the star radius. This illustrates
 the effect of rotation on the minimal baryonic mass. 
The parameters of the EOS chosen
are $s_{\rm env}=5$, $s_{\rm core}=1~{\rm k_{B}}$, 
$n_{\rm env} = 0.02~{\rm fm^{-3}}$ 
and $n_{\rm core} = 0.26~{\rm fm}^{-3}$, with $Y_l=0.4$ in the core and
$Y_\nu=0$ in the shocked envelope.
\label{fig:statmass}}
\end{figure}
As one can see, there is a minimal mass for such equilibria, as for 
cold ones, but it is of the order of 
${M}^{\rm min}_{\rm bar}({\rm EPNS})
 \sim 1.4~ {\rm M}_\odot$, much higher than the standard 
minimal mass
of cold NS (${M}^{\rm min}_{\rm bar}({\rm NS})
 \sim 0.1~ {\rm M}_\odot$). 
This high minimal mass of EPNS 
is due to the fact that thermal pressure at the base of the neutrinosphere
 tends to blow-off the envelope of the star, and thus additional
mass compared to the cold case is needed to preserve equilibrium.
This effect was already present in the models of paper I, but we were
at that time interested in the maximal mass of PNS and did not investigate
this problem. For late-type PNS, the minimal mass 
is increased at most up to 
${M}^{\rm min}_{\rm bar}({\rm LPNS})
 \sim 0.9~ {\rm M}_\odot$ (Gondek et al., 1997). 
\begin{table*}
\centering
\caption[]{
Minimal baryonic masses of EPNS for various values of the parameters of
the EOS.
}
\begin{tabular}{llllllll}
\hline
 &&&&&&&\\
$s_{\rm env}$                   & 5    & 4    & 5    & 5    & 5    & 5\\
$s_{\rm core}$                  & 1    & 1    & 2    & 1    & 1    & 1\\ 
$n_{\rm env}$~ [${\rm fm^{-3}}$]& 0.02 & 0.02 & 0.02 & 0.02 & 0.02 & 0.01\\
$n_{\rm core} [{\rm fm^{-3}}$]   & 0.26 & 0.26 & 0.26 & 0.3  & 0.1  & 0.26\\
&&&&&&&\\
\hline
&&&&&&&\\
${M}_{\rm bar,min}$ [${\rm M}_\odot$] &1.41  & 1.23 & 1.56 & 1.50 & 1.19 & 1.22\\
&&&&&&&\\
\hline
\end{tabular}
\end{table*}

The value of ${M}_{\rm bar}^{\rm min}({\rm EPNS})$ is strongly linked to the
properties of our EOS, through the assumed values of the 
EOS parameters. Table 1 displays the minimal masses
reachable for various combinations of the parameters of the EOS.

For some sets of parameters, minimal baryonic mass is larger than 
$1.5~{\rm M}_\odot$.
In particular, we found that for $s_{\rm env} \ge 6$, the minimal baryonic 
mass is greater than $1.5~{\rm M}_\odot$ and therefore 
no static equilibrium can be found for our  ``canonical''  baryon mass.
Let us notice, that the value of 
minimal mass is not sensitive to the presence of trapped neutrinos in the 
unshocked core : had we taken a neutrino free core,
the minimal mass would change only by a few percent as compared 
to that calculated assuming $Y_l=0.4$. 

Let us stress, that EPNS models with $n_{\rm centr}$ lower than 
that corresponding to the minimum mass configuration (i.e., those 
to the left of the minima of the  mass - central baryon density 
curves in Fig. 3) are rather unrealistic, and moreover likely to 
be unstable. They consist mostly of the shocked, 
  high entropy matter. Clearly, such  objects could not 
be formed in a successful SN II explosion: too much of the shock energy 
would be dissipated for heating and dissociating the collapsing matter.
  
The very existence of  a high minimal baryonic mass for static 
configurations of  early-type
PNS leads to  severe constraints on their rotational velocity or, more
precisely, on the maximal angular momentum they can bear.
As one increases the central angular velocity of such stars, keeping
their baryonic mass constant, the
centrifugal force adds to the thermal pressure at the base of the
envelope. The envelope is then more easily lifted than in the static
case, leading to an increase of the minimal mass 
of stationary configurations. At the ``minimal mass'' limit, the
minimal mass coincides with
the canonical one and no equilibrium can be found with
a higher angular momentum.
 So, apart from
the usual keplerian - or ``mass shedding'' - limit (for which the 
centrifugal force exactly balances
the gravity at the equator), this ``minimal mass''
limit also constrains the domains of variations of $\Omega_c$ and $R_0^2$.
%
\subsection{Variations of $J$ versus $\Omega_c$ for early-type PNS}
Given the two limits constraining $\Omega_c$ and $R_0^2$, the problem
remains of finding the maximal angular momentum of EPNS.
Variations of the angular momentum versus central rotational velocity 
were previously studied for cold uniformly rotating NS with various
realistic EOS (Cook et al. 1992, 1994, Salgado et al. 1994).
These studies showed that there exist two distinct families of uniformly rotating
cold NS : {\it normal} and {\it supra-massive} stars.
Normal stars have a baryonic mass lower than the maximal static baryonic
mass, and their angular momentum is an increasing function of the central
angular velocity. Supra-massive stars have  baryonic mass greater than
the maximal static baryonic mass, which they would thus not support
without rotation. Their angular momentum is not a monotonic function
of the central angular velocity, and the stable keplerian configuration
is not that of maximal angular momentum. This property
leads so to the well-known phenomenon of ``spin-up by angular momentum
loss''.

\begin{figure}     
\centerline{
\epsfig{figure=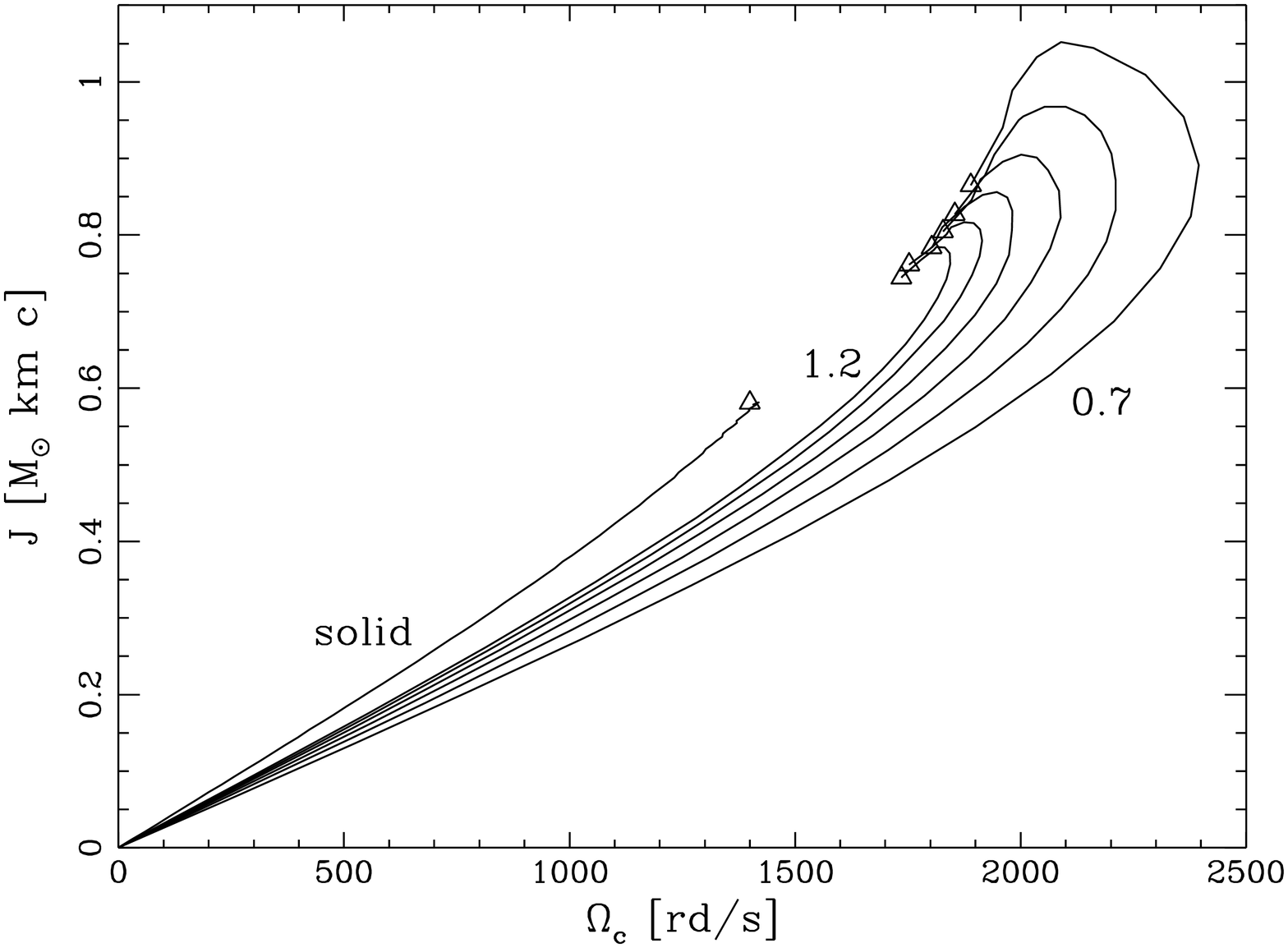,width=8.5cm,angle=-90}}
\caption[]{{\bf a.}
Angular momentum versus central angular velocity of EPNS  
of baryon mass $1.5~{\rm M}_\odot$, for 
rigid rotation and for differential rotation with $R_0^2 = 1.2$, 1.1,
 1.0, 0.9, 0.8 and 0.7 km$^2$ (from the left to the right).
The parameters of the EOS are as in Fig. \ref{fig:statmass}. 
Triangles denote the keplerian configurations. 
\label{fig:jomeg1}}
\addtocounter{figure}{-1}
\end{figure}
\begin{figure}     
\centerline{
\epsfig{figure=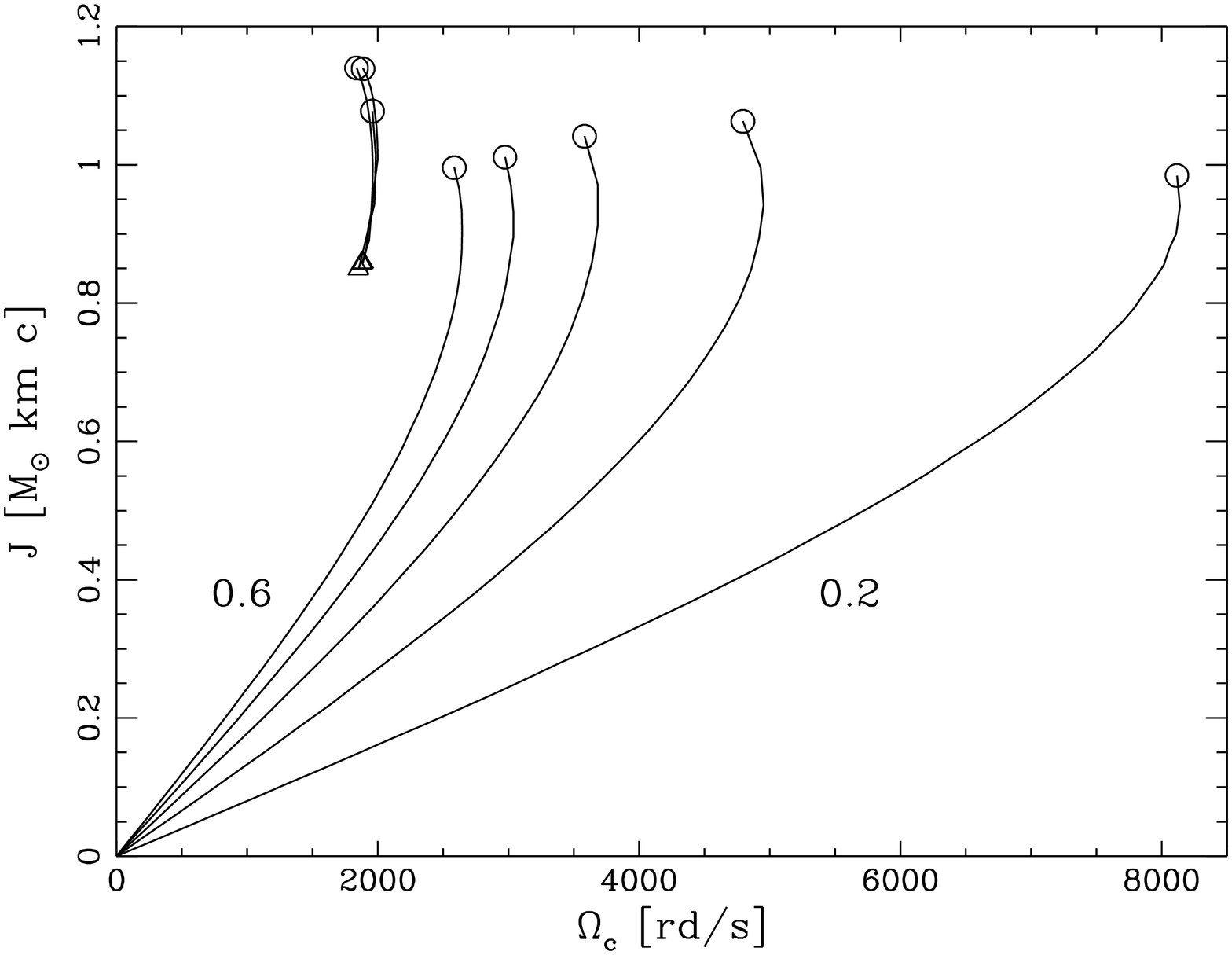,width=8.5cm,angle=-90}}
\caption[]{{\bf b.}
Angular momentum versus central angular velocity of early-type PNS, 
of baryon mass $1.5~{\rm M}_\odot$,  
for differential rotation with $R_0^2 = 0.6$, 0.5,
 0.4, 0.3 and 0.2 km$^2$ (from the left to the right).
The parameters of the EOS are as in Fig. \ref{fig:statmass}. 
Triangles denote the keplerian configurations and circles are for
configurations with $M_{\rm bar}^{\rm min}=1.5~{\rm M}_\odot$ 
(``minimal mass'' limit configurations). The upper-left curves are
the $J_{\rm ms}<J<J_{\rm mm}$ parts of the $R_0^2 = 0.6$, 0.5,
 0.4 km$^2$ curves (see text).
\label{fig:jomeg2}}
\end{figure}
As the canonical mass we chose ($M_{\rm bar}=1.5~{\rm M}_\odot$) is less
than the maximal static baryonic mass for all the EOS we used,
we expected that, for a given $\Omega$-contrast (a given $R_0^2$),
 $J$ would increase with increasing central angular
velocity for our models of PNS. Numerical experiments showed that 
this is indeed the case for cold models but {\it not}
 for the hot ones. This can be understood as follows.
For cold uniformly rotating NS, the phenomenon of spin-up by angular
 momentum loss is due to the high oblateness of the supra-massive stars due
to their high rotational velocities (see
Cook et al., 1992, 1994) : since $\Omega = J/I$,  the variations of $\Omega$
depend on the variations of both $J$ and $I$, for a sufficiently extended
star, the redistribution of $I$ when 
one increases $\Omega$ can lead
to a decreasing $J$. This effect is thus strongly linked to the
radial distribution of matter inside the star, and can be significantly
enhanced by the presence of differential rotation (Shapiro et al., 1990),
but nevertheless appears {\it only} in supra-massives stars.
This is not the case for PNS. Due to their extended
structure induced by the supplementary thermal pressure, very little rotation 
is needed to produce this phenomenon. Thus it can appear even in 
{\it normal} PNS.

Figures \ref{fig:jomeg1}a and \ref{fig:jomeg1}b depict
 typical behaviours of $J$ as a function
of $\Omega_c$ : starting from the static model,
we increase the quantity $R \Omega_c$ (which parametrizes our models)
up to the limiting configurations, keeping the baryonic mass and $R_0^2$ 
constant. As one can see, the ``spin-up'' phenomenon is not
present for uniformly rotating shocked PNS. Differential rotation has the
effect of increasing the maximal central angular velocity 
and the maximal angular momentum $J^{\rm max}$, ``pushing'' the curves 
to the right and the top in Fig. \ref{fig:jomeg1}. It also increases
the angular momentum $J_{\rm ms}$ at which mass shedding appears (configurations
denoted by triangles on Fig. \ref{fig:jomeg1}a) : $J_{\rm ms}$ is an increasing
function of $1/R_0^2$.

The two physical limits
previously mentioned restrain the available maximal angular momentum.
For small $\Omega$-contrast (large $R_0^2$), the mass shedding limit
is reached first, and one cannot store enough angular momentum in
the star to reach the ``minimal mass'' limit at $J=J_{\rm mm}$ : this is 
the case for all the configurations of Fig. \ref{fig:jomeg1}a, for which 
 the  angular 
momentum  $J$ verifies $0 \leq J < J^{\rm max}$, with   
$J_{\rm ms}<J^{\rm max}<J_{\rm mm}$. 
Let us stress that in the case of cold uniformly rotating normal NS, one 
has always $J^{\rm max} = J_{\rm ms}$ for a fixed $M_{\rm bar}$, which is
never the case 
for the differentially rotating sequences displayed in 
Fig. 4a, b. 

As we decrease $R_0^2$, $J^{\rm max}$ increases and 
the $J^{\rm max}$ configuration will finally reach 
the minimal
mass limit for $J^{\rm max}=J_{\rm mm} 
\sim 1 ~{\rm M}_\odot~{\rm km~c}$ for the EOS
we chose. With a further increase of the $\Omega$-contrast 
(decrease of $R_0^2$), the $J-\Omega_{\rm c}$ plot at fixed 
$M_{\rm bar}$ divides into two disconnected segments (branches). 
The first branch, which starts at $\Omega_{\rm c}=0$, ends at the minimum mass 
limit with $J_{\rm mm}$. A second segment 
(upper left curves in Fig. \ref{fig:jomeg1}b) is bounded from above 
by another minimum mass limit, $J'_{\rm mm}$ (which is actually an 
overall maximum of $J$ for a specific value of $R_0^2$), 
and from below by the mass shedding limit 
with $J_{\rm ms}$. Notice, that these segments evolved  from the upper 
left parts of the $J$--$\Omega_{\rm c}$ curves in 
Fig. \ref{fig:jomeg1}a, with the very upper part removed by the minimum mass 
constraint. The upper left segments correspond to rotating configurations 
 with  $J_{\rm ms} \leq J \leq J'_{\rm mm}=J^{\rm max}$. 
Such a situation is characteristic of a rather narrow range of the 
$R_0^2$ parameter 
 ($R_0^2=$ 0.6, 0.5 and 0.4 km$^2$ in 
Fig. \ref{fig:jomeg1}b). For a still larger $\Omega$-contrast 
(still smaller $R_0^2$) the left segment disappears 
 and we are left only with the 
configurations verifying $0 \leq J \leq J_{\rm mm}=J^{\rm max}$ 
(curves with $R_0^2=$ 0.3, 0.2 km$^2$ in 
Fig. \ref{fig:jomeg1}b).

We show in Fig. \ref{fig:profilkep}a and \ref{fig:profilmass}b the 
 constant energy density lines in  the cross sections of 
two typical limiting configurations of  
EPNS. As one can see, the keplerian configuration is a very extended
one, with a ``disk-like'' shape due to the combined effect of 
differential rotation and of a high entropy envelope. The minimal mass limit
configuration is more compact, with a ``toroidal'' shape. 
In both cases, small irregularities
in the outermost (low density)  constant energy density lines  
are due to the Gibbs phenomenon
in the $\theta$ direction, a problem inherent in our numerical 
scheme but small enough not to affect badly the virial 
parameter (see sect. 4.1).
\begin{figure}     
\centerline{
\epsfig{figure=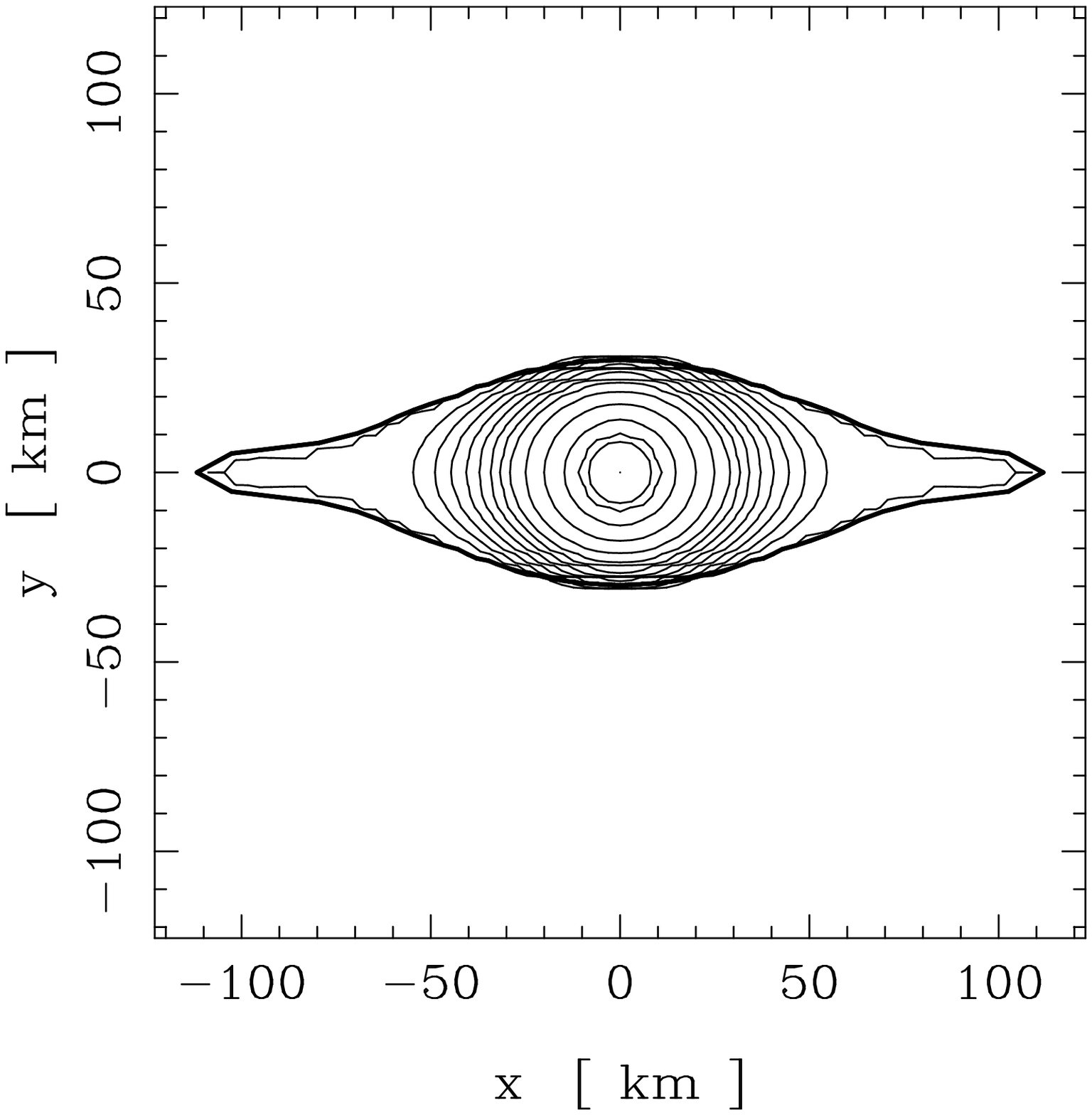,width=7.cm,angle=-90}}
\caption[]{{\bf a.}
Iso-energy density lines of the keplerian configuration of 
$M_{\rm bar}=1.5~{\rm M}_\odot$, with $R_0^2=0.5$ km$^2$ (corresponding to 
the triangle of the $R_0^2=0.5$ curve in figure \ref{fig:jomeg2}b). $x$ and 
$y$ are related to the 
MSQI coordinates $r$ and $\theta$ by $x=r\cos\theta$, $y=r\sin\theta$
(see BGSM for details).
\label{fig:profilkep}}
\addtocounter{figure}{-1}
\end{figure}
\begin{figure}     
\centerline{
\epsfig{figure=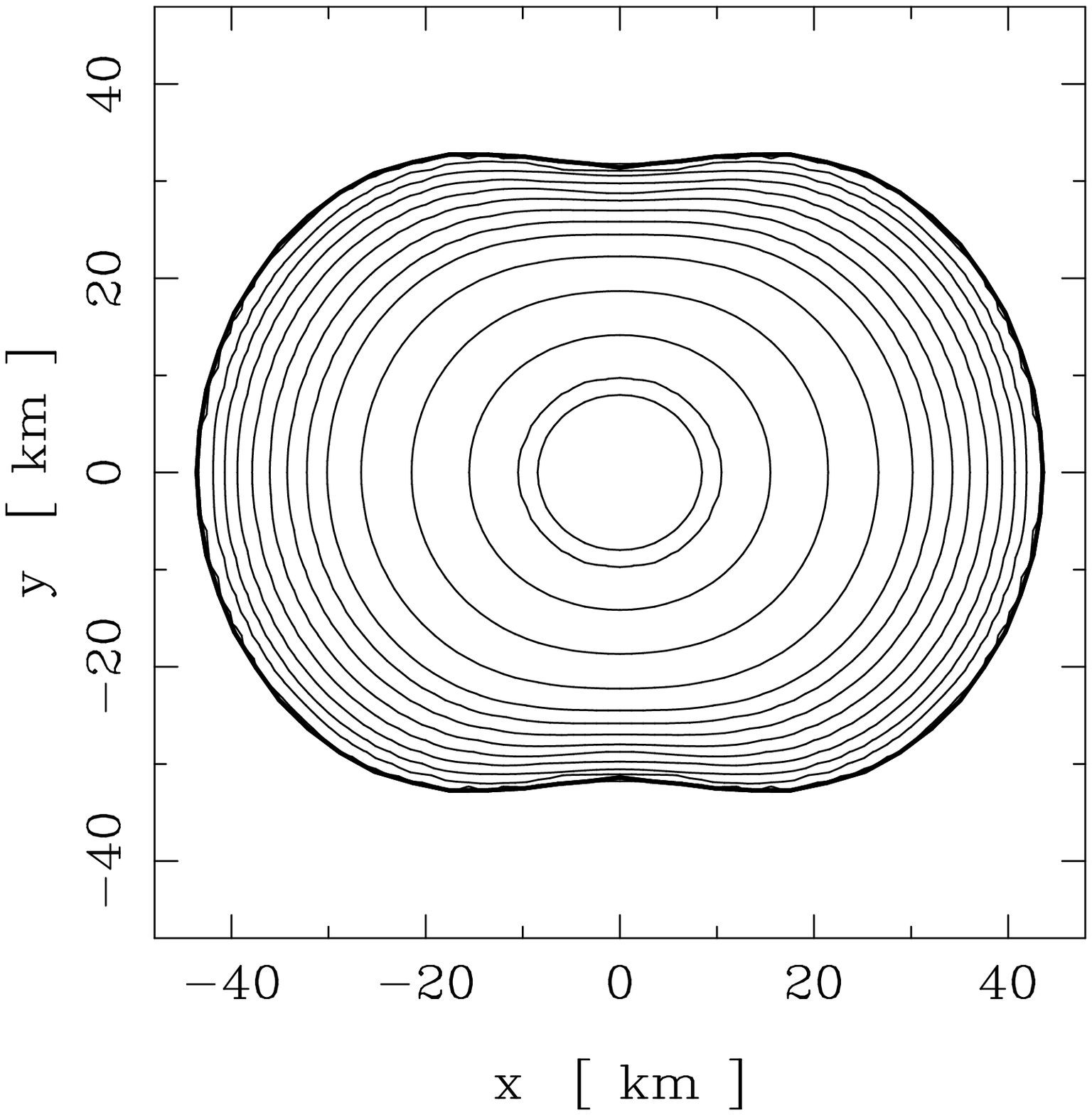,width=6.8cm,angle=-90}}
\caption[]{{\bf b.}
Iso-energy density lines of the minimal mass limit configuration of 
$M_{\rm bar}=1.5~{\rm M}_\odot$, with $R_0^2=0.5$ km$^2$ (corresponding to 
the upper right circle of the $R_0^2=0.5$ curve in figure \ref{fig:jomeg2}b).
\label{fig:profilmass}}
\end{figure}
%
\subsection{Thermal structure and maximal angular momentum of PNS}
Uniformly rotating models of late-type PNS, considered in Paper I, assumed a 
relatively homogeneous distribution of entropy within the 
neutrino opaque interior. In order to answer the question, to 
what extent the conclusions of Paper I would be modified by allowing 
for differential rotation, we considered models, which were isentropic 
or isothermal within the neutrino opaque core.  Let us mention, that such 
a homogeneous  thermal state of a PNS corresponds to a relatively late 
 switching-off of accretion due to the lift-off of the shock. Accretion stage 
had to be sufficiently long to allow for settling-down of the PNS envelope, 
and for  the thermal uniformization of the neutrino-opaque interior. In 
what follows, we will restrict ourselves to the case of 
$M_{\rm bar}=1.5~{\rm M}_\odot$. As in the case of two-component EPNS,
 described 
in the preceding section, we calculated $J(\Omega_c)$ sequences for fixed 
parameter $R_0^2$. 
We found, that there exist a specific value, $R_{\rm 0,crit}$, such that 
for $R_0<R_{\rm 0,crit}$ we get the maximum angular momentum of PNS which is 
larger than the maximum angular momentum  of cold, uniformly rotating NS of 
the same $M_{\rm bar}$. So, for $R_0<R_{\rm 0,crit}$, differential rotation 
allows PNS to have enough angular momentum to yield cold NS, rotating 
uniformly at maximal angular velocity. The values of $R_{\rm 0,crit}$ for the
different EOS used in paper I are displayed in table 2.
\begin{table*}
\caption[]
{Parameter $R_{\rm 0,crit}$ for different states of the LPNS 
interior}
\begin{tabular}{ccccc}
\hline
&&&&\\
EOS  &$s_{\rm core}=2$&$s_{\rm core}=1$&$T^\star=15$ MeV&$T^\star=25$ MeV\\
     &$Y_l=0.4$       &$Y_l=0.4$       &$Y_\nu=0$       &$Y_\nu=0$ \\
&&&&\\
\hline
&&&&\\
$R_{\rm 0,crit}^2~[{\rm km}^2]$ & 1.044 & 1.091 & 0.916 & 0.894 \\
&&&&\\
\hline
\end{tabular}
\end{table*}
It should be stressed, that this 
qualitative conclusion is generally valid for thermally homogeneous LPNS, 
for all considered EOS of stellar interior (isentropic with $s\le 2$, 
isothermal, with no trapped lepton number $Y_\nu=0$, 
 and  with maximum trapped lepton fraction $Y_l=0.4$), and that the 
actual value of $R_{\rm 0,crit}$ is roughly independent of the assumed EOS.

For $R_0<R_{\rm 0,crit}$,  
  we can consider configurations 
 having an angular momentum greater than the maximal
angular momentum of cold NS. Such PNS  will necessarily eject a part of $J$ and
$M_{\rm bar}$ 
 during their evolution toward cold, uniformly rotating  NS. 
This will probably lead to a  maximally rotating  (i.e., rotating at its 
mass shedding limit) NS surrounded 
by an ejected matter. For $R_0>R_{\rm 0,crit}$, evolution of PNS into 
a cold NS leads to the final rotation frequency which is lower than 
the maximal one for such a cold NS, but the difference decreases with 
decreasing $R_0$.   

Let us consider now  the case of early-type PNS. 
 Being aware of the particular behaviour of $J$ depicted in the previous
paragraph, we scanned the whole domain of $0.1 < R_0^2 < 1~~{\rm km}^2$
for the maximal angular momentum of LPNS. In this domain,
the maximal angular momentum configurations were never keplerian.
We found that $ 1.0 < J_{\rm max} < 1.16~~{\rm M}_\odot {\rm km~c}$, which 
is roughly independent of $R_0^2$. 
As the crucial
factor determining the ``minimal mass limit'' is the effective gravity at 
the base of the shocked envelope for large $\Omega$-contrast, it
is not surprising to find that our results are roughly {\it independent} of
the parameter $R_0^2$ (thus, in some way, of the assumed rotation law), 
and depends {\it only} on the angular
momentum of the star. These limits on the total angular momentum of the 
early-type PNS leads to limits on the period of cold NS of the same
baryon mass of $1.6 < P < 1.8$ ms. 

These results are relevant for NS which originate
in a type II supernova and evolve {\it without significant accretion} from 
a hot, differentially rotating LPNS to a cold,  uniformly rotating
NS. If the baryonic mass of such a PNS is $\sim 1.5~{\rm M}_\odot$, then the
maximum frequency of the resulting NS cannot exceed $\sim 600$ Hz. 
This evolutionary
argument thus imposes more stringent limits on the rotation rate of NS than
standard calculations (Cook et al. 1994, 
Salgado et al. 1994).
%
\section{Discussion and conclusions}
In the present paper we studied models of rapidly differentially 
rotating protoneutron stars. Using realistic equations of state 
of the protoneutron star interior, we calculated, using general 
relativistic equations of stationary motion, families of configurations 
corresponding to the main stages of evolution of protoneutron stars. 
The earliest stage corresponds to  two-component objects, with a 
low entropy unshocked core and a bloated, high-entropy envelope. 
The next stage is that of a lepton rich protoneutron star, 
with a relatively uniform distribution of entropy per baryon within  
the hot neutrino-opaque core, containing degenerate neutrinos. 
Finally, the latest stage is that a deleptonized hot neutrino-opaque 
core, which contains non-degenerate neutrinos. We considered mostly 
 the case of a protoneutron star of the ``canonical baryon mass'' 
 $1.5~{\rm M}_\odot$, which roughly corresponds to measured 
gravitational masses of binary pulsars. 

Configurations of stationary motion were calculated assuming axial 
symmetry, and neglecting effects of meridional circulation and 
convection. Such an approximation was valid, because the velocities 
of meridional circulation, as well as those connected with 
convective motion, were much smaller than the velocity of sound 
in the protoneutron star interior. On the other hand, they were also 
much smaller than the linear equatorial velocities considered. 
 While convective motions could be assumed to have no direct 
dynamical effect, they influenced the PNS structure through 
changes in composition and temperature,  through transport of 
heat and  of lepton number, leading to the evolution of the 
equation of state of stellar interior. 
  Our assumption that the local temperature 
was a function of local density  was sufficient  for the 
equations of stationary motion to be integrable; stationary state  
of differential rotation was then that of a barotropic equilibrium. 

The approximation of stationarity was crucial for the simplicity of 
calculations. The shear viscosity, resulting from neutrino diffusion, 
implies angular momentum transport and heating  of the stellar 
interior. However, the estimates of the characteristic timescales, 
under various regimes prevailing within the hot neutrino opaque 
core, imply that these effects can be neglected on the timescale 
of the PNS lifetime. Clearly, the timescale of 
angular momentum and energy transport can actually be significantly 
shortened by convective and turbulent motions within the stellar 
interior. Also, the presence of magnetic field, and subsequent 
 amplifying of it due to the differential character of rotation
(winding-up of magnetic field lines),  
could contribute to the angular momentum transport and 
to rigidifying of the rotational motion. However, even in 
this case characteristic timescales referring to the global 
structure of rotating star may be  expected to be of the 
order of seconds. Clearly, all these effects which are expected 
to be operating in a differentially rotating protoneutron star, 
deserve a separate study.

Our models describe rotating PNS after revival and lift-off of the 
 shock, 
i.e., after accretion onto PNS became insignificant for mechanical 
equilibrium. Basically, we considered two limiting  scenarios of 
the `liberation' of PNS from accretion. The first case corresponds 
to a relatively early switching-off of accretion, so that rotating 
PNS has still a pronounced two-component thermal structure, with an 
unshocked low entropy core and a shocked high entropy envelope. The 
second case is that of a relatively  late lifting-off of the shock, 
when the accretion stage was sufficiently long to allow for mixing 
of the low and high entropy matter, leading to a roughly isentropic 
PNS interior. Let us notice, however, that the second scenario is, 
in view of very high accretion rate, quite risky for the PNS survival. 
Huge accretion rates (of the order of ${\rm M}_\odot/$s) might result in 
a high probability of exceeding the maximum allowable mass for neutron 
stars, and therefore could easily lead to formation of a black hole. 

For a ``canonical baryon mass'' of $1.5~{\rm M}_\odot$, and within the 
considered family of differentially rotating configurations, 
 we find a set of configurations with angular momentum exceeding 
the maximum one that can be accommodated by rigidly rotating cold 
neutron stars of the same baryon mass. We expect, that the evolution 
of such differentially rotating protoneutron stars into cold, 
uniformly rotating neutron stars will be accompanied by ejection 
of matter from their equators. Such a mass shedding configuration 
may be expected to be susceptible to some additional instabilities, 
which - if breaking axial symmetry - could contribute to a further 
angular momentum loss via emission of gravitational radiation. 

Assuming that the initial differentially rotating configuration 
had a characteristic two-component structure, with a high entropy 
envelope, containing some half of the baryon mass of the protoneutron 
star, we get (for a standard baryon mass of $1.5~{\rm M}_\odot$) 
 a minimum period of  resulting cold neutron star  
$P_{\rm min}^{\rm cold}\simeq 1.7~$ms. Shorter rotation periods 
of cold neutron star 
of the baryon mass of $\sim 1.5~{\rm M}_\odot$ 
  (such as the periods  measured for the most 
rapid millisecond pulsars like PSR 0034-0534, PSR 1937+21,
 PSR 1957+20, see e.g. Bailes \& Lorimer 1995)  could then 
be reached only by a spin-up via accretion of 
a significant fraction of solar mass from an accretion disk. 
\begin{acknowledgements} 
We are very grateful to J-P. Zahn and S. Talon for introducing us to the
difficult subject of instabilities in differentially rotating fluids. 
This research was partially supported by the JUMELAGE program 
``Astronomie France-Pologne'' of CNRS/PAN and by the KBN grant No. P304 014 07.
The numerical computations have been 
performed on Silicon Graphics workstations, purchased thanks to the 
support of the SPM department of the CNRS and the Institut National des 
Sciences de l'Univers.  

\end{acknowledgements}
%

\end{document}